 \magnification=1200
\input amssym.def
\input amssym.tex
 \font\newrm =cmr10 at 24pt
\def\bul{\raise .9pt\hbox{\newrm .\kern-.105em } }

 \def\fr{\frak}
 \font\sevenrm=cmr7 at 7pt

 \baselineskip 16pt

 \def\h{\hbox{ }}

 \def\n{{\fr n}}
 \def\a{{\fr a}}

 \def\b{{\fr b}}
 
 \def\hh{{\fr h}}

 \def\g{{\fr g}}

 \def\<{\le}
 \def\>{\ge}

 \def\s{{\h\subset\h}}

 \def\vs{\vskip }

 \def\mapright#1
  {\smash{\mathop
  {\longrightarrow}
  \limits^{#1}}}

 \def\kk#1{{\kern .4 em} #1}
 \def\vs{\vskip 1pc}

\font\titlefont=cmr10 at 14pt
\font\authorfont=cmr10 at 12pt
\font\ninerm=cmr10 at 9pt
\font\ninebf=cmb10 at 9pt
\font\sevenrm=cmr10 at 7pt

\rm

 \centerline{\titlefont Experimental evidence for the occurrence of  E$_8$ in
nature}\vskip 2pt \centerline{\titlefont  and the radii of the
Gosset circles}\vskip 1.5pc \centerline{\authorfont Bertram
Kostant}\vskip 1.5pc \noindent \baselineskip11pt  {\ninebf
Abstract}. \ninerm A recent experimental discovery involving the
spin structure of electrons in a cold one-dimensional magnet points
to a validation of a Zamolodchikov model involving the exceptional
Lie group E$_8$. The model predicts 8  particles and predicts the
ratio of their masses. In more detail, the vertices of the
8-dimensional Gosset polytope identifies with the 240 roots of
E$_8$. Under the famous two-dimensional (Peter McMullen) projection
of the polytope,  the image of the vertices are arranged in 8
concentric circles, hereafter referred to as the Gosset circles. The
Gosset circles are now understood to correspond to the 8 masses in
the model,  and in addition it is understood that the ratio of the
their radii is the same as the ratio of the corresponding
conjectural masses. A ratio of the two smallest circles (read 2
smallest masses) is the golden number. Marvelously,  the conjectures
have been now validated experimentally, at least for the first five
masses.

 The McMullen projection generalizes to any complex simple Lie algebra whose rank is
greater than 1. The Gosset circles generalize as well, using orbits
of the Coxeter element. Using results in [K-59], I  found
some time ago a very easily defined operator $A$  whose spectrum is
exactly the squares of the radii $r_i$ of  these generalized Gosset
circles. As a confirmation, in the E$_8$ case, using only the
eigenvalues of a suitable multiple of $A$, Vogan  computed the ratio
of the $r_i$. Happily these agree with the corresponding ratio of
the Zamolodchikov masses.

 The operator $A$ is written as a sum of $\ell +1$ rank 1
operators, parameterized by the points in the extended Dynkin diagram. Involved
in this expansion are the coefficients
$n_i$ of the highest root. Suggestively, recalling the McKay correspondence, in the
 E$_8$ case the $n_i$, together with 1,  are the dimensions of the irreducible
representations of the binary icosahedral group.

\vskip 4pt \noindent {\ninebf Key Words:} E$_8$, Cartan subalgebras in apposition,
Gosset circles, Ising chain in E$_8$ symmetry, Zamolodchikov theory,
1-dimensional magnet, Coxeter element, Coxeter number,  golden
number, conformal field theory, particle physics. \vskip 4pt
\noindent {\ninebf Mathematics Subject Classification (2010):}
Primary 20G41, 20G45, 81R05, 81T10.

 \vskip 1.5pc \baselineskip =13pt  \centerline{\bf 0.
Introduction}\vskip 1.5pc  {\bf 0.1.} \rm  Let $\g$ be a complex simple Lie algebra and let
$(x,y)$ be the Killing form ${\cal B}$ on $\g$. Let
$\ell=\hbox{rank}\,\g$ and let $\hh$ be a Cartan subalgebra of $\g$. Let
$\Delta$ be the set of roots for $(\hh,\g)$ and let $\Delta_+\s
\Delta$ be a choice of positive roots. For any $\varphi \in
\Delta$, let $e_{\varphi}$ be a corresponding root vector. We
assume choices are made so that $(e_{\varphi},e_{-\varphi}) = 1$.

Let
$\Pi =
\{\alpha_1,\ldots,\alpha_{\ell}\}\s \Delta_+$ be the set of simple
positive roots. Let $h$ be the Coxeter number of $\g$. Let
$w\in\hh$ be the unique element such that $\langle \alpha_i,w\rangle =1,\,
i=1,\ldots,\ell$.  Let $G$ be a Lie group such that
$\hbox{Lie}\,G = \g$ and let $H\s G$ be the subgroup corresponding to
$\hh$. Let $c\in
H$ be given by $c = \hbox{exp}\,2\pi i\,w/h$. Then $c$ is a regular semisimple element of $G$ and its
centralizer in $\g$ is given by $$\g^c = \hh.\eqno (0.1)$$\vskip .5pc

Let $\psi\in \Delta_+ $ be the highest root and let $n_i,\,i=1,\ldots, \ell $, be the coefficients (known to
 be positive) relative to the simple roots so that
$\psi=\sum_{i=1}^{\ell} n_i\,\alpha_i$. Let $$x(\beta) = e_{-\psi} + \sum_{i=1}^{\ell} \,\sqrt n_i
e_{\alpha_i}.\eqno (0.2)$$ Then results in [K-59] imply
$x(\beta)$ is a regular semisimple element of $\g$. Let $\hh(\beta)$ be the Cartan subalgebra of
  $\g$ which contains $x(\beta)$. Furthermore
let $\gamma= e^{2\pi i/h}$  so that $\gamma $ is a primitive $h$ root of unity.  The following is proved
in [K-59].\vs

{\bf Theorem 0.1} {\it The Cartan subalgebra $\hh(\beta)$ is stable under ${\hbox{\rm Ad}}\,c$. Furthermore if
$\sigma_{\beta} = \hbox{\rm Ad}\,c|\hh(\beta)$, then $\sigma_{\beta}$ is a Coxeter element in the Weyl group $W(\beta)$
of
$\hh(\beta)$. In addition $$\sigma_{\beta}\,x(\beta) = \gamma\,x(\beta).\eqno (0.3)$$}\vskip.5pc

The two
Cartan subalgebras $\hh$ and $\hh(\beta)$ are said to be in apposition, in the terminology of [K-59].

Let $\Delta(\beta)$ be the set of roots for the pair $(\hh(\beta, \g)$. Then $\Delta(\beta)$ decomposes into $\ell$ orbits,
$O_i,\,i=1,\ldots, \ell$, under the
 action of $\sigma(\beta)$ and each orbit $O_i$ has $h$ roots. We may choose root vectors $e_{\nu}$ for $\nu\in \Delta$ so that one has $c\cdot e_{\nu}
 = e_{\sigma_{\beta}\nu}$ for any $\nu\in \Delta(\beta)$. \vs

 {\bf Theorem 0.2.} {\it The elements $z_i,
\,i=1,\ldots,\ell$ in $\g$, given by $$z_i = {1\over
h}\,\sum_{\nu\in O_i} e_{\nu}$$ are a basis of $\hh$.}\vs

{\bf 0.2.} We assume from now on $\ell>1$
so that $h>2$. Let $\hbox{Vec}\,\hh(\beta)$ be the  real space of all hyperbolic elements in $\hh(\beta)$ so that
$\hbox{Vec}\,\hh(\beta)$ is a $W(\beta)$-stable real form of $\hh(\beta)$. If conjugation in $\hh(\beta)$ is
defined with respect to $\hbox{Vec}\,\hh(\beta)$,  then $\overline{x(\beta)}$ is a regular eigenvector of
$\sigma_{\beta}$ with eigenvalue $\overline {\gamma}$. Since $\gamma\notin \Bbb R$ one defines a
real two-dimensional $\sigma_{\beta}$-stable subspace $Y$ of $\hbox{Vec}\,\hh(\beta)$ by putting
$$Y = \hbox{Vec}\,\hh(\beta)\cap(\Bbb C x(\beta) +\Bbb C \overline {x(\beta)}).\eqno (0.4)$$
 Let $$Q: \hbox{Vec}\,\hh(\beta)\to Y\eqno (0.5)$$ be the orthogonal projection
(and $\sigma_{\beta}$ map) defined by (positive definite) ${\cal B}| \hbox{Vec}\,\hh(\beta)$. For any $\nu\in \Delta$
 let $w_{\nu}\in \hbox{Vec}\,\hh(\beta)$ be the image of
$\nu$ under the $W(\beta)$-isomorphism $\hh(\beta)^*\to \hh(\beta)$ defined by ${\cal B}$.
One defines circles $C_i,\,i=1,\ldots,\ell,$ in $Y$ of positive radius
$r_i$, centered at the origin, by the condition that $$Q(w_{\nu})\in C_i,\,\,\forall \nu\in O_i.$$
In the special case where $G=E_8$ we will refer to the
$C_i$ as Gosset circles. \vs

{\bf Remark.} If $\gamma'$ is another primitive $h$ root of unity, then one knows
(A.J. Coleman) that $\gamma'$ occurs with multiplicity 1 as an eigenvalue of $\sigma_{\beta}$. If $0\neq
x(\beta)'$ is a corresponding (necessarily regular) eigenvector, then one may replace $(\gamma, x(\beta))$ by
$(\gamma', x(\beta)')$ and replace $(Q,Y)$ by a corresponding $(Q',Y')$. However both cases are
``geometrically"  isomorphic. In particular the radii $r_i$ do not change. The reason for this is that one can
show that if $Z(\sigma(\beta))$ is the cyclic group generated by $\sigma(\beta)$ and $N(\sigma(\beta))$ is the
normalizer of $Z(\sigma(\beta)$ in $W(\beta)$, then $$N(\sigma(\beta))/Z(\sigma(\beta))\cong \Gamma_h, \eqno
(0.6)$$
 where $\Gamma_h$ is the Galois group of the
cyclotomic field spanned over $\Bbb Q$ by the $h$ roots of unity.

In the E$_8$ case the projection $Q$ of the Gosset polytope appears
ubiquitously throughout the mathematical literature  (see e.g., the
frontispiece of [CX]). It has been described by Coxeter as the
``most symmetric" two-dimensional projection of this polyhedron. But
in fact it can be described precisely as the unique such projection,
up to isomorphism, (there are 4 such projections) which commute with
the action of the Coxeter element $\sigma_{\beta}$.

The following Theorem 0.3 below  is our main result. It gives the
radii $r_i$ of the circles $C_i$. Its significance in the E$_8$ case
will be explained in \S 0.3. If $x,y\in \hh$, let $x\otimes y$ be
the rank 1 operator on $\hh$, defined so that if $z\in \hh$, then
$x\otimes y(z) = (x,z)\,y$. Also let $w_i\in\hh$ be the image of
$\alpha_i$ in $\hh$ under the isomorphism $\hh^*\to\hh$ defined by
the Killing form ${\cal B}$.
 Now let $A$ be the operator on $\hh$, written as a sum
of $\ell+1$ rank 1 operators, given by putting $$A = \sum_{i=0}^{\ell } n_i\,w_i \otimes w_i, \eqno (0.7)$$
 where $n_0 = 1$ and, we recall, $n_i,\,i>0, $ is the
coefficient of $\alpha_i$ in the simple root expansion of the highest root $\psi$.\vs

 {\bf Theorem 0.3.} {\it The eigenvalues of $A$ are $r_i^2,\, i=1,\ldots,\ell,$
and $z_i$ (see Theorem 0.2) is an $A$-eigenvector for $r_i^2$. }

\vs {\bf 0.3.} I obtained the result Theorem 0.3 sometime (I
believe) in the early 1990s. At that time publication seemed
unwarranted since I believed there would be little interest in a
knowledge of the radii $r_i$.  However in the middle 1990s Peter
McMullen's image, by $Q$, in the E$_8$ case, of the Gosset polytope
was very widely published and became well known, even to many in the
general public, due no doubt, to the very extensive (and well
deserved) publicity, given to the determination by a large team of
mathematicians, of the characters of the real forms of E$_8$. I
showed Theorem 0.3 to David Vogan, one of the leading members of the
aforementioned team. Applying a computer program to a Weyl group
reformulation of a scalar multiple of the operator $A$, the
following list, in increasing size, of the normalized  8 radii was
obtained by Vogan. His normalization was to make the largest of the
8 radii equal to 1. To avoid decimals we took the liberty of
multiplying his list by 1000. (What will be significant is the ratio
of the radii and not the radii themselves).
$$\matrix{209\cr 338\cr 416\cr 502\cr 618\cr 673\cr 813\cr 1000\cr}\eqno (0.8)$$\vskip .2pc {\bf Remark.}
The first 7 numbers in (0.8) are the integral parts of
the normalized radii and to that extent only approximate the normalized radii.  \vskip 8pt

Having been directed by colleagues to the papers [Za] and [Co], we have only recently become aware of the fact that the
ratio of the numbers in (0.8) have physical significance. In [Za] Zamolodchikov conjectures the existence  of 8 particles in
connection with a field theory associated with the Ising model. Happily his ratio of the masses of the conjectured particles
``agrees" with the ratio of the normalized radii in (0.8). (See (1.8), p.~4237 in [Za].) The use of the quotation marks in
``agrees", and also below, is because of the statement in the remark above. Particular emphasis is made  in [Za] of the fact
that the ratio ($m_2/m_1$) of the first two masses should be the golden number.
 Indeed $338/209\, ``=" {1\over 2}(1+ \sqrt
5)$. Zamolodchikov makes a connection in his paper with E$_8$ at the
bottom of p.~4247 but attributes the prediction of an E$_8$
connection to V.~Fateev.

The recent nine person authored paper [Co] is an experimental discovery, using a very
cold one-dimensional magnet, validating Zamolodchikov's theory, at least for the first 5 particles. In particular the equality of
$m_2/m_1$ (the ratio of the radii of the two inner Gosset circles) with the golden number is very clearly seen. \vs
{\bf 0.4.} I wish to thank Nolan Wallach for many conversations regarding the subject matter in this paper. In particular for conversations regarding the two types
of Gosset circles which appear in a number of publications. In addition I thank for him for the minicourse
he gave me on the representations of the Virasoro
algebra. Representations of this algebra make an appearance in [Za]. I also want to thank David Vogan for
making the computations, using Theorem 0.3, which resulted in (0.8). Also David factorized a relevant
 characteristic polynomial into a product of two irreducible (over $\Bbb Q$) polynomials of
degree 4. The irreducible polynomials relate directly to the two types of Gosset circles mentioned above.
\vskip 1pc

\centerline{\bf 1. Cartan subalgebras in
apposition}\vskip 1pc {\bf 1.1.} Let $\g$ be a complex simple Lie algebra and let
$(x,y)$ be the Killing form ${\cal B}$ on $\g$. Let
$\ell=\hbox{rank}\,\g$ and let $\hh$ be a Cartan subalgebra of $\g$. Let
$\Delta$ be the set of roots for $(\hh,\g)$ and let $\Delta_+\s
\Delta$ be a choice of positive roots. For any $\varphi\in
\Delta$ let $e_{\varphi}$ be a corresponding root vector. We
assume choices are made so that $$(e_{\varphi},e_{-\varphi}) =
1. \eqno (1.1)$$
Let
$\Pi =
\{\alpha_1,\ldots,\alpha_{\ell}\}\s \Delta_+$ be the set of simple
positive roots. Let $h$ be the Coxeter number of $\g$. Let
$w\in\hh$ be the unique element such that for all
$i=1,\ldots,\ell,$ $$\langle \alpha_i,w\rangle =1. \eqno (1.2)$$
For any $\varphi\in \Delta$ put $o(\varphi) = \langle
\varphi,w\rangle$. If $\psi\in \Delta_+$ is the highest root, then one knows that
$$o(\psi) = h-1. \eqno (1.3)$$
Let $G$ be a Lie group such that
$\hbox{\rm Lie}\,G = \g$ and let $H\s G$ be the subgroup corresponding to
$\hh$. If $a\in H$ and $\nu\in \hh^*$ is an $H$-weight, put
$a^{\nu} = e^{\langle x,\nu\rangle}$, where $a = \hbox{exp}\,x$. Let $c\in
H$ be given by $$c = \hbox{exp}\,2\pi i\,w/h. \eqno (1.4)$$ Also let
$$\gamma = e^{2\pi i/h}.$$ For $g\in G$ and $z\in \g$
we will sometimes write $g\cdot z$ for $\hbox{Ad}\,g(z)$. Let
$$\g(\gamma) =
\{x\in
\g\mid c\cdot x = \gamma\,x\}.$$ As one readily notes, using e.g., (1.3), \vs

 {\bf
Proposition 1.1.} {\it One has $\hbox{\rm dim}\,\g(\gamma) = \ell +1$ and
in fact the elements $e_{\alpha_i},\,i =1,\ldots,\ell,$  and
$e_{-\psi}$ are a basis of $\g(\gamma)$.}\vs

For $$\beta =
(\beta_1,\ldots,\beta_{\ell}, \beta_{-\psi}) \in \Bbb C^{\ell
+1}$$ let $x(\beta)\in \g(\gamma) $ be defined by putting
$$x(\beta) =
\beta_{-\psi}\,e_{-\psi}+ \sum_{i=1}^{\ell}
\beta_i\,e_{\alpha_i}.\eqno (1.5)$$ The following result was
established in [K-59]. \vs

{\bf Theorem 1.2.}
{\it $x(\beta)$ is regular semisimple if and only if $\beta\in
(\Bbb
C^{\times})^{\ell +1}$. } \vs

If $\beta \in (\Bbb
C^{\times})^{\ell +1}$, let $\hh(\beta)$ be the Cartan subalgebra
which contains $x(\beta)$. It is immediate that if $\beta\in (\Bbb
C^{\times})^{\ell +1}$, then $\hh(\beta)$ is stable under $\hbox{Ad}\,c$.
Let $\sigma_{\beta}$ be the element of the Weyl group of
$\hh(\beta)$ defined by $c$. One thus has $$\sigma_{\beta}
\,x(\beta) = \gamma\,x(\beta).\eqno (1.6)$$ Part of the following is
established in [K-59] and uses a result of A.J. Coleman. \vs

{\bf
Theorem 1.3.} {\it Let $\beta\in (\Bbb C^{\times})^{\ell +1}$. Then
$\sigma_{\beta}$ is a Coxeter element of the Weyl group of
$\hh(\beta)$ and up to a scalar multiple $x(\beta)$ is the unique
element of $\hh(\beta)$ satisfying (1.6).}\vs

{\bf 1.2.} Using the Killing form we may identify the algebra of polynomial
functions on $\g$ with the symmetric algebra $S(\g)$. This is done
so that if $x,y\in \g$, then $x^n(y) = (x,y)^n$. Also the Killing
form extends naturally to a symmetric bilinear form $(p,q)$ on
$S(\g)$. One has $${1 \over n!}\, (x^n,y^n) = x^n(y).$$ The algebra of
$S(\g)^G$ of symmetric invariants is a polynomial ring $\Bbb
C[J_1,\ldots, J_{\ell}]$ where the $J_k$ are homogeneous, say of
degree $d_k$, and algebraically independent. Choose the ordering
so that the $d_k$ are nonincreasing. In that case $d_1 = h$ and
$d_k<h$ for $k>1$. The definition of a cyclic element $x\in \g$
was introduced in [K-59]. One has that $x$ is cyclic if and only if
$J_1(x)\neq 0$ and $J_k(x) = 0$ for $k>1$. This condition is
independent of the choice of the $J_k$. It is established in
[K-59] that cyclic elements are regular semisimple. We have also
proved \vs

 {\bf Theorem 1.4.} {\it $x(\beta)$ is cyclic for any
$\beta\in (\Bbb C^{\times})^{\ell +1}$ and, up to conjugacy, any cyclic element in
$\g$ is of this form.} \vs

Let $n_i\in \Bbb C, i=1,\ldots,\ell,$ be
defined so that $$\psi = \sum_{i=1}^{\ell} n_i \, \alpha_i.$$ One
knows that the $n_i$ are positive integers. It is immediate from
the independence of the simple roots that if $k_i\in \Bbb
Z_+,\,i=1,\ldots,\ell$, and
$k_{-\psi}\in \Bbb Z_+$ are such that $k_{-\psi} +
\sum_{i=1}^{\ell}k_i = h$,  then the monomial
$$e_{-\psi}^{k_{-\psi}}\,e_{\alpha_1}^{k_1}\cdots
e_{\alpha_{\ell}}^{k_{\ell}}\eqno (1.7)$$ is in $S^h(\g)$ and

\vs
{\bf Proposition 1.5.} {\it The monomial (1.7) is a zero weight vector if and only if
$k_{-\psi} =1$ and $k_i = n_i$ for $i=1,\ldots,\ell$.} \vs

It
follows immediately from Theorem 1.4 and Proposition 1.5 that if
$\beta\in  (\Bbb C^{\times})^{l+1}$, then

 \vs {\bf Proposition 1.6.}
{\it With respect to the inner product in S(\g) one
has $$(J_1,e_{-\psi}\, e_{\alpha_1}^{m_1}\cdots
e_{-\alpha_{\ell}}^{m_{\ell}})\neq 0$$ and
$$J_1(x(\beta)) =
\beta_{-\psi}\,\beta_1^{m_1}\cdots\beta_{\ell}^{m_{\ell}}\,{1\over
m_1!\cdots m_{\ell}!}(J_1,e_{-\psi}\, e_{\alpha_1}^{m_1}\cdots
e_{-\alpha_{\ell}}^{m_{\ell}}).\eqno (1.8)$$}

 \vs It is clear that
the set of cyclic elelemnts of the form  $x(\beta)$ for $\beta\in
(\Bbb C^{\times})^{l+1}$ is stable under conjugation by $H$. One
readily defines an action of
$H$ on
$(\Bbb C^{\times})^{l+1}$ so that if
$\beta\in (\Bbb C^{\times})^{l+1}$ and $a\in H$, then $$ a\cdot
x(\beta) = x(a\cdot \beta).$$ Of course $(\Bbb C^{\times})^{l+1}$ is
stable under multiplication by $\Bbb C^{\times}$. One defines another
action of $\Bbb C^{\times}$ on $(\Bbb C^{\times})^{l+1}$ where, for $\lambda\in \Bbb
C^{\times}$ and $\beta \in (\Bbb C^{\times})^{l+1}$, one lets
$\lambda\ast \beta \in (\Bbb C^{\times})^{l+1}$ be given so that
$$(\lambda\ast \beta )_i = \beta_i, \,i=1,\ldots, \ell, $$ but
$$(\lambda\ast \beta )_{-\psi} = \lambda\,\beta_{-\psi}.$$ \vskip .5pc {\bf
Theorem 1.7.}

{\it (1) Two cyclic elements $v,v'$ in
$\g$ are
$G$-conjugate if and only if $J_1(v)= J_1(v')$.

 (2) Furthermore,
if $v=x(\beta),\,v'=  x(\beta')$ where $\beta,\beta\in (\Bbb
C^{\times})^{l+1}$,  then $v$ and $v'$ are $G$-conjugate $\iff$ they are
$H$-conjugate.

(3) Given a cyclic element $v\in \g$ and $\beta\in (\Bbb C^{\times})^{l+1}$,  there
exists a unique $\lambda\in \Bbb C^{\times}$ such that $v$ and $ x(\lambda\ast \beta)$
are $G$-conjugate.}

 \vs {\bf Proof.} (1) follows immediately from the fact that cyclic
elements are regular semisimple. To prove (2), assume that $x(\beta)$ and $x(\beta')$
are
$G$-conjugate. Clearly there exists $a\in H$ so that $(a\cdot \beta')_i=\beta_i$ for
$i=1,\ldots,\ell$. But then $(a\cdot \beta')_{-\psi}=\beta_{-\psi}$ by (1.8) since
$J_1(x(\beta)) = J_1(a\cdot \beta')$. Thus $a\cdot x(\beta') = x(\beta)$. (3)
follows from (1) since $J_1(x_{\lambda\ast \beta})$ is linear in $\lambda$ by (1.8).
\hfill QED

\vs
{\bf 1.2.} Let
$G_{\hbox{\sevenrm Ad}}$ be the adjoint group. Then we recall from [K-59] there
exists a unique conjugacy class
$C$ of regular elements of order $h$ in $G_{\hbox{\sevenrm Ad}}$. Furthermore if
$a\in G$ and $\hbox{Ad}\,a\in C$, there exists a Cartan subalgebra $\a$
which is stable under
$\hbox{Ad}\,a$ and
$\hbox{Ad}\,a\mid \a$ is a Coxeter element. In such a case we will say
that $\a$ is Coxeter for $a$. Conversely, if $\sigma$ is a
Coxeter element for a Cartan subalgebra $\a$ and $\hbox{Ad}\,a$ normalizes
$\a$ and induces $\sigma$, then $\hbox{Ad}\,a\in C$.

Recalling Theorem 1.3, one has that
$c\in C$. Furthermore since any Cartan subalgebra which is Coxeter for $c$
necessarily has a regular eigenvector with eigenvalue $\gamma$ for $\hbox{Ad}\,c$, one has \vs

{\bf Proposition 1.8.} {\it $\hh(\beta)$ is Coxeter for $c$ for any $\beta \in
(\Bbb C^{\times})^{l+1}$. Conversely, any Cartan which is Coxeter for $c$ is equal
to
$\hh(\beta)$ for some $\beta\in (\Bbb C^{\times})^{l+1}$. }\vs

Now note that by
Theorem 1.3, if
$\beta,\beta'\in (\Bbb C^{\times})^{l+1}$, then $$\hh(\beta) = \hh(\beta')\,\,\iff
\beta' = \lambda\,\beta\,\,\hbox{for some}\,\,\lambda\in \Bbb C^{\times}. \eqno
(1.9)$$ On the other hand, if $a\in H$ and $\beta \in (\Bbb C^{\times})^{l+1}$, then
obviously $$\hbox{Ad}\,a\,\,(\hh(\beta)) = \hh(a\cdot \beta).\eqno (1.10)$$\vskip .5pc

 {\bf
Theorem 1.9.} {\it  Let $c'\in C$. Then the set of all Cartan subalgebras which are
Coxeter for $c'$ is an adjoint orbit for the (unique) Cartan subgroup which contains
$c'$. In particular if $c=c'$, then $H$ is the Cartan subgroup which contains $c$ and
the orbit is\break $\{\hh(\beta)\mid \beta\in (\Bbb C^{\times})^{l+1}\}$.}

 \vs {\bf Proof.} Let
$\beta, \beta'\in (\Bbb C^{\times})^{l+1}$, We have only to show that there exists
$a\in H$ such that $$\hbox{Ad}\,a\,\,(\hh(\beta)) = \hh(\beta').\eqno (1.11)$$ But now for any
$\lambda\in \Bbb C^{\times}$ one has $$J_1(x(\lambda\,\beta')) = \lambda^h\,
J_1(x(\beta')).\eqno (1.12)$$ But then we can choose $\lambda$ so that
$$J_1(x(\lambda\,\beta')) = J_1(x(\beta)).\eqno (1.13)$$ But then by Theorem 1.7 there
exists $a\in H$ such that $a\cdot x(\beta) = x(\lambda\,\beta')$. But then clearly
$\hbox{Ad}\,a\,(\hh(\beta)) = \hh(\lambda\,\beta')$. But $\hh(\lambda\,\beta')
=\hh(\beta')$. \hfill QED

\vs {\bf 1.3.} Let $\beta\in (\Bbb C^{\times})^{l+1}$ and let
$\Delta(\beta)$ be the set of roots for the pair $(\hh(\beta),\g)$. Then
$\sigma_{\beta} = \hbox{Ad}\,c|\hh(\beta)$ is a Coxeter element.  Let
$O_i\s
\Delta(\beta), i=1,\ldots,\ell,$ be the orbits of $\sigma_{\beta}$. For any $\nu\in
\Delta(\beta)$, let $e_{\nu}\in \hh(\beta)^{\perp}$ be a corresponding
root vector. We assume the root vectors are chosen so that $$c\cdot e_{\nu} =
e_{\sigma_{\beta}\,\nu}.$$ We note, for $\nu \in O_i$, that we may write $$e_{\nu} =
z_i +
\sum_{\varphi\in \Delta}d_{\nu,\varphi}\, e_{\varphi},$$ where $z_i\in \hh$. One further notes
that $$\eqalign{c\cdot e_{\nu} &= z_i +
\sum_{\varphi\in \Delta}\gamma^{o(\varphi)}\,d_{\nu,\varphi}\, e_{\varphi}\cr
&= e_{\sigma_{\beta}\nu}\cr &= z_i + \sum_{\varphi\in
\Delta}\,d_{\sigma_{\beta}\nu,\varphi}
\,e_{\varphi}.\cr}\eqno (1.14)$$  However since $c$ is regular one has $$\g^c =\hh.$$
But
$\sum_{\nu\in C_i}\,e_{\nu}$ is an invariant of
$c$ and hence lies in $\hh$. Thus from (1.14) one must have $$\sum_{\nu\in
C_i}\,e_{\nu} = h\,z_i.\eqno (1.15)$$ That is, for any $\varphi\in \Delta$,
$$\sum_{\nu\in C_i}\,d_{\nu,\varphi} = 0,\eqno (1.16)$$ and in fact the orbits $O_i$
consequently define a distinguished basis of $\hh$.

\vs {\bf Theorem 1.10}. {\it The $z_i,\, i=1,\ldots,\ell$, given by (1.15), are a
basis of $\hh$.} \vs {\bf Proof.} Since $\sigma_{\beta}$ has no nontrivial invariant
in $\hh$ the only contribution to $\g^c$ must come from (1.15). But this proves the
theorem since there are $\ell$ orbits and $\hbox{dim}\,\hh = \ell$.

\hfill  QED

\vs {\bf 1.4.} Now
note that (1.1) implies
$$[e_{\varphi},e_{-\varphi}] = w_{\varphi}\eqno (1.17)$$ for any $\varphi\in \Delta$,
where $w_{\varphi}\in \hh$ is such that for any $x\in \hh$, $$(w_{\varphi},x) =
\langle\varphi,x\rangle. \eqno (1.18)$$ Now recalling the notation of (1.8) one has
by (1.18),  $$\sum_{i=1}^{\ell} n_i\,w_{\alpha_i} = w_{\psi}.\eqno
(1.19)$$ Now any Cartan subalgebra $\hh_1$ is the sum of its vector part $\hbox{Vec}\, \hh_1$
(split real Cartan subalgebra) and its toroidal part $\hbox{Tor}\,\hh_1= i\,\hbox{Vec}\, \hh_1 $
(Cartan subalgebra of a compact real form). In particular, for any
$\beta\in (\Bbb C^{\times})^{\ell +1}$ one has $$\hh(\beta) = \hbox{Vec}\,\hh(\beta) +
i\,\hbox{Vec}\,
\hh(\beta).\eqno (1.20)$$ In particular $$x(\beta) = \Re\,x(\beta) +
i\Im\,x(\beta),\eqno (1.21)$$ where $\Re\, x(\beta),\,\Im\,x(\beta)\in \hbox{Vec}\,
\hh(\beta)$. But now recalling (1.6) one has
$$\eqalign{\gamma\,x(\beta) &= (\Re\,\gamma + i\,\Im\,\gamma)(\Re\,x(\beta) +
i\Im\,x(\beta))\cr &= (\Re\,\gamma\,\Re\,x(\beta)-\Im\,\gamma\,\Im\,x(\beta)) +
i\,(\Re\,\gamma\, \Im\,x(\beta) + \Im\,\gamma\, \Re\,x(\beta) ).\cr} \eqno (1.22)$$
Since $c_{\beta}$ stabilizes both vector and toroidal parts of $\hh(\beta)$ one has
$$\eqalign{c_{\beta}\,\Re\, x(\beta)&=
\Re\,\gamma\,\Re\,x(\beta)-\Im\,\gamma\,\Im\,x(\beta)\cr c_{\beta}\,\Im\,
x(\beta)&= \Re\,\gamma\, \Im\,x(\beta) + \Im\,\gamma\, \Re\,x(\beta). \cr}\eqno
(1.23)$$ Now put $$\overline {x(\beta)} = \Re\,x(\beta) - i\,\Im\,x(\beta)\eqno
(1.24)$$ so that $\overline {x(\beta)}\in \hh(\beta)$. Also, for
 $\nu\in \Delta(\beta)$,
let $$\nu_{\beta} = \langle \nu, x(\beta)\rangle. \eqno (1.25)$$ \vskip .5pc
 One notes that $$\eqalign{\sigma_{\beta}\overline{x(\beta)}& =
\overline{\gamma}\,\overline{x(\beta)}\cr \langle \nu, \overline{x(\beta)}\rangle &=
\overline{\nu_{\beta}}.\cr}\eqno (1.26)$$

 \vs Indeed the first equation in
(1.26) is obvious from (1.23). The second follows from the fact that $\nu$ takes real
values on $\hbox{Vec}\,\hh(\beta)$.

\vskip .5pc Now let
$$x_-(\beta) = {1\over
\beta_{-\psi}}\, e_{\psi} +
\sum_{i=1}^{\ell}\,{n_i\over \beta_i}\, e_{-\alpha_i}.\eqno (1.27)$$\vskip .5pc It
is immediate from (1.3) that $$c\cdot x_-(\beta) = \overline {\gamma}\,
x_-(\beta).\eqno (1.28)$$

\vs {\bf Theorem 1.11.} {\it One has $x_-(\beta)\in
\hh_{\beta}$ and $$\sigma_{\beta}  x_-(\beta) =  \overline {\gamma}\,
x_-(\beta).\eqno (1.29)$$ In fact by Coleman's uniqueness theorem there exists
$t_{\beta}\in \Bbb C^{\times}$ such that $$t_{\beta}\, \overline {x(\beta)} =
x_-(\beta).\eqno (1.30)$$}

\vs {\bf Proof.} Since $x(\beta)\in \hh(\beta)$ and is
regular it clearly suffices to prove that $x_-(\beta)$ commutes with $x(\beta)$. Here
one recalls (1.26) and (1.28). But
$$\eqalign{[x(\beta),x_-(\beta)]&= [\beta_{-\psi}\,e_{-\psi}+ \sum_{i=1}^{\ell}
\beta_i\,e_{\alpha_i},{1\over
\beta_{-\psi}}\, e_{\psi} +
\sum_{i=1}^{\ell}\,{n_i\over \beta_i}\, e_{-\alpha_i}]\cr &=
[e_{-\psi},e_{\psi}] +
\sum_{i=1}^{\ell}\,n_i [e_{\alpha_i},e_{-\alpha_i}]\cr &= w_{-\psi} +
\sum_{i=1}n_i\,w_{\alpha_i}\cr& = 0\,\,\hbox{by (1.19)}. \qquad \qquad\qquad \qquad\qquad \qquad\qquad \qquad\qquad
\qquad\hbox{QED}\cr} $$

 \vs Normalize
(Weyl's normal form) the choice of the $e_{\varphi},\,\varphi\in \Delta$ so that
$$\theta (e_{\varphi}) = - e_{-\varphi},\eqno(1.31)$$ where $\theta$ is an involution
of
$\g$ such that $\theta = -1$ on $\hh$. In particular there exists a compact form $\g_u$
of $\g$
$$e_{\varphi}-e_{-\varphi}\,\,\hbox{is contained in $\g_u$ for all $\varphi\in
\Delta$.} \eqno (1.32)$$ Let
$\beta^{(1)}\in (\Bbb C^{\times})^{l+1}$ be defined so that $\beta^{(1)}_{-\psi}= 1$
and $\beta^{(1)}_i = \sqrt {n_i},\,i=1,\ldots,\ell$. Then one has

\vs {\bf Theorem
1.12.} {\it One has $x_-(\beta^{(1)}) = \overline{x(\beta^{(1)})}$ so that
$t_{\beta^{(1)}} = 1$}. \vs {\bf Proof.} One has $$x(\beta^{(1)}) = e_{-\psi} +
\sum_{i=1}\sqrt {n_i} e_{\alpha_i}\eqno(1.33)$$ and $$x_-(\beta{(1)}) = e_{\psi} +
\sum_{i=1}\sqrt {n_i} e_{-\alpha_i}.\eqno(1.34)$$ But then
$$x(\beta^{(1)})-x_-(\beta^{(1)})= e_{-\psi} -e_{-\psi} + \sum_{i=1}^{\ell}\sqrt
{n_i}(e_{\alpha_i}-e_{-\alpha_i}).\eqno (1.35)$$ But then
$x(\beta^{(1)})-x_-(\beta^{(1)})\in \g_u$ by (1.32) and hence in particular, the
corresponding operator, for the adjoint representation, has pure
imaginary spectrum. Thus $x(\beta^{(1)})-x_-(\beta^{(1)})\in \Im\,\hh(\beta^{(1)})$.
But then $$\overline {x(\beta^{(1)})} - x_-(\beta^{(1)})\in \Im\,\hh(\beta^{(1)})$$
Hence $$(t_{\beta^{(1)}}-1)(x_-(\beta^{(1)}))\in \Im\,\hh(\beta^{(1)})
$$ by (1.30). That is, if $s = i (t_{\beta^{(1)}}-1)$, then $$s
(x_-(\beta^{(1)}))\in
\Re \,\hh(\beta^{(1)})$$ On the other hand $\Re \,\hh(\beta^{(1)})$ is stable under
$\sigma_{\beta^{(1)}}$. But this implies that $s=0$ since otherwise one has the
contradiction that $s\, (x_-(\beta^{(1)}))$ is a eigenvector for
$\sigma_{\beta^{(1)}}$ with eigenvalue $\overline {\gamma}$ by (1.29). Hence
$t_{\beta^{(1)}}=1$. The theorem then follows from (1.30). \hfill QED

\vs {\bf 1.5.} We
recall the notation of the first paragraph of \S 1.3. Let $\b$ be the Borel
subalgebra defined by $\Delta_+$ and let $\n$ be the nilradical of $\b$. Let
$
\beta\in (\Bbb C^{\times})^{l+1}$ and let $\beta'\in (\Bbb C^{\times})^{\ell}$ be
defined by deleting the last entry $\beta_{-\psi}$ from $\beta$. Let $x(\beta') =
x(\beta)- \beta_{-\psi}e_{-\psi}$ so that $x(\beta')\in \n$ is principal
nilpotent. For the opposed direction let $\overline {\b} = \theta\,\b$ and let
$\overline {\n}= \theta\,\n$. Then let $x_-(\beta') = x_-(\beta)- {1\over
\beta_{\psi}} e_{\psi}$ so that $x_-(\beta')$ is principal nilpotent in
$\overline{\n}$. Let
$i\in \{1,\ldots,\ell\}$ and let $\nu\in O_i$. We will now see that the root vector
$e_{\nu}$ for $\hh(\beta)$ is completely determined by its component $z_i\in \hh$ and
the number $\nu_{\beta} = \langle \nu, x(\beta)\rangle$ (see (1.25). Recall that the
regularity of $x(\beta)$ guarantees that $$\nu_{\beta}\neq 0.\eqno (1.36)$$ For $k\in
\Bbb Z$ let
$$\g(k) =
\{x\in\g\mid [w,x] = k\,x\}$$ so that one has the direct sum $$\g =\oplus_{k=
-\infty}^{\infty}\g(k),\eqno (1.37)$$ and let $$P_k:\g\to \g(k)$$ be the projection
defined by (1.36). We will let
$e_{\nu}(k) = P_k\,e_{\nu}$ so that $$e_{\nu} = \sum_{k= -\infty}^{\infty}
e_{\nu}(k), \eqno (1.38)$$ noting that $$e_{\nu}(0) = z_i.\eqno (1.39)$$ Of course
$\g(k) = 0$ for $|k|\geq h$ so that $e_{\nu}(k) = 0$ for $|k|\geq h$. Also $\g(0) =
\hh$ and $e_{\nu}(0) = z_i$.

\vs {\bf Theorem 1.13.}
{\it Let
$\beta\in (\Bbb C^{\times})^{l+1}$ and $\nu \in O_i$. Then for any positive integer
$k$ one has
$$\eqalign{e_{\nu}(k) &= {1\over \nu_{\beta}}\, [x(\beta'),e_{\nu}(k-1)]\cr &=
{1\over \nu_{\beta}^k}\, (\hbox{\rm ad}\,x(\beta'))^k\,z_i.\cr}\eqno (1.40)$$}

\vs {\bf Proof.} By
induction we have only to prove the first line of (1.40). But now from the root
vector property $$e_{\nu} = {1\over \nu_{\beta}}[x(\beta), e_{\nu}].\eqno (1.41)$$ But
since $e_{\nu}(j) = 0$ for $j\geq h$ the only contribution to $e_{\nu}(k)$ on the left
side of (1.41) is where $x(\beta')$ replaces $x(\beta)$ on the right side of (1.41)
and $e_{\nu}(k-1)$ replaces $e_{\nu}$. \hfill QED

\vs {\bf Remark.} If one considers
the principle TDS defined by ${1\over \nu_{\beta}}\,x(\beta')=e'$ and $w$, then note
that Theorem 1.12 asserts that $e_{\nu}(k)$ for $k>0$ are just the elements in
the cyclic $e'$-module generated by $z_i$.
\vs But we can also reverse direction. By (1.26) and (1.30)
$$\eqalign{\langle \nu, x_-(\beta)\rangle &= t_{\beta}\, \langle \nu, \overline
{x(\beta)}\rangle \cr &= t_{\beta}\,\overline{\nu_{\beta}}.\cr}$$ But then
 $e_{\nu}$
is an eigenvector for $\hbox{\rm ad}\, (t_{\beta}\,\overline{\nu_{\beta}})^{-1}\, x_-(\beta)$
with eigenvalue 1.  An argument similar  to that in the proof of Theorem 1.12
yields

\vs {\bf Theorem 1.14.} {\it Let the notation be as in Theorem 1.12 and (1.30).
Then for any positive integer
$k$ one has
$$\eqalign{e_{\nu}(-k) &= [(t_{\beta}\,\overline{\nu_{\beta}})^{-1}\,
x_-(\beta'),e_{\nu}(-k + 1)]\cr &= ((ad\,(t_{\beta}\,\overline{\nu_{\beta}})^{-1}\,
x_-(\beta'))^k z_i. \cr}\eqno(1.42)$$ }

 {\bf Lemma 1.15.} {\it Let $z\in \hh$
and
$\beta\in (\Bbb C^{\times})^{l+1}$. Then the maximum positive integer $M_z$ such that
$(\hbox{\rm ad}\,x(\beta'))^{M_z} \,z\neq 0$ is independent of $\beta$. Moreover $M_z$ is the
maximum integer such that $(\hbox{\rm ad}\,x_-(\beta'))^{M_z} \,z\neq 0$. In fact if $\a$ is
any principal TDS containing $w$, then $2M_z +1$ is the dimension of the
maximal dimensional irreducible component of the $\hbox{\rm ad}\,\a$ submodule generated by
$z$}.

\vs {\bf Proof.} Immediate from observation that any two such principal TDS are
 conjugate under the action of $s\hbox{Ad}\,\hbox{exp}\,\hh$. \hfill QED

\vs Recalling the notation and
statements of Theorems 1.12 and 1.14, put $M_i = M_{z_i}$ so that $$e_{\nu}(k) =
0\,\,\hbox{for $|k| > M_i$.}\eqno (1.43)$$ Now we may write $e_{\nu}(h-1) =
r_{\nu}e_{\psi}$ for some scalar $r_{\nu}$. Also by
(1.42) $$\eqalign{[x(\beta'),e_{\nu}(-1)]&= [x(\beta'),[
(t_{\beta}\,\overline{\nu_{\beta}})^{-1}\, x_-(\beta'),z_i]]\cr &=
(t_{\beta}\,\overline{\nu_{\beta}})^{-1}\,
[x(\beta'),[x_-(\beta'),z_i]]\cr
&=(t_{\beta}\,\overline{\nu_{\beta}})^{-1}\sum_{j=1}^{\ell} \langle
\alpha_j,z_i\rangle n_j w_{\alpha_j}\cr}\eqno (1.44)$$ so that
$$\eqalign{z_i &= {1\over
\nu_{\beta}}([x(\beta'),e_{\nu}(-1)] +
\beta_{-\psi}\,r_{\nu} [e_{-\psi},e_{\psi}])\cr   &= {1\over
\nu_{\beta}}((t_{\beta}\,\overline{\nu_{\beta}})^{-1}\,\sum_{j=1}^{\ell} \langle
\alpha_j,z_i\rangle n_j w_{\alpha_j} - \beta_{-\psi}\,r_{\nu} w_{\psi})\cr &={1\over
\nu_{\beta}}(\sum_{j=1}^{\ell}((t_{\beta}\,\overline{\nu_{\beta}})^{-1}\langle
\alpha_j,z_i\rangle -
\beta_{-\psi}\,r_{\nu}) n_j w_{\alpha_j}).\cr}\eqno (1.45)$$

Next one recalls that since $$[x_-(\beta),e_{\nu} ] = t_{\beta}[\overline
{x(\beta)},\,e_{\nu}],\eqno (1.46)$$ computing the component in $\g_{h-1}$, one has
$${1\over \beta_{-\psi}}\,[e_{\psi}, z_i] = t_{\beta}\overline
{\nu_{\beta}}\,r_{\nu}e_{\psi}\eqno (1.47)$$ so that $${\langle -\psi, z_i\rangle\over
\beta_{-\psi}}= t_{\beta}\overline {\nu_{\beta}}\,r_{\nu}.\eqno (1.48)$$
That is, $$-\beta_{-\psi}\,r_{\nu} = (t_{\beta}\,\overline{\nu_{\beta}})^{-1}\langle
\psi,z_i\rangle  $$ so that
$${|\nu_{\beta}|^2\over t_{\beta}}\,z_i =
\sum_{j=1}^{\ell}\langle
\alpha_j + \psi,z_i\rangle\,n_j w_{\alpha_j}.\eqno (1.49)$$\vskip .5pc

{\bf 1.6.} We will identify $\hbox{End}\,\hh$ with $\hh\otimes \hh$, where if $x,y \in \hh$,
then $x\otimes y\in \hbox{End}\,\hh$ is that operator such that if $z\in \hh$, then $$x\otimes
y (z) = (x,z)\,y$$ Then if $A\in \hbox{End}\,\hh$ is given by $$A = \sum_{j=1}^{\ell}
(w_{\alpha_j} + w_{\psi}) \otimes n_jw_{\alpha_j}$$, then for $z\in \hh$
one has $$A\,z = \sum_{\j=1}^{\ell} \langle \alpha_j + \psi, z\rangle
n_j\,w_{\alpha_j}.$$ Then (1.49) is the statement

\vs {\bf Proposition
1.16.} {\it For $i=1,\ldots,\ell,$ one has that $z_i$ is an eigenvector of $A$ with
eigenvalue $$|\nu_{\beta}|^2/t_{\beta}.\eqno (1.50)$$}\vs We now want to simplify the
expression for $A$. Indeed $$\eqalign{ \sum_{j=1}^{\ell}
(w_{\alpha_j} + w_{\psi}) \otimes n_jw_{\alpha_j}&= \sum_{j=1}^{\ell}
w_{\alpha_j} \otimes n_jw_{\alpha_j} + \sum_{j=1}^{\ell}
 w_{\psi} \otimes n_jw_{\alpha_j}\cr&= \sum_{j=1}^{\ell}
(w_{\alpha_j} \otimes n_jw_{\alpha_j}) + w_{\psi}\otimes
(\sum_{j=1}^{\ell}n_j\,w_{\alpha_j})\cr &= \sum_{j=1}^{\ell}n_j
(w_{\alpha_j} \otimes w_{\alpha_j}) + w_{\psi}\otimes w_{\psi}.\cr}\eqno (1.51)$$ Thus
if we consider the extended Dynkin diagram adding another node $\alpha_0 = -\psi$ and
define $m_0 =1$ as in the McKay correspondence, we have proved

 \vs {\bf Theorem
1.17.} {\it One has $$A = \sum_{j=0}^{\ell} n_j\,\, w_{\alpha_j}\otimes
w_{\alpha_j}.\eqno (1.52)$$}

 {\bf 1.7.} Henceforth we fix $\beta$ so that $\beta = \beta^{(1)}$ (see
Theorem 1.11) so that $t_{\beta} = 1$. Also assume $\g$ is not of type $A_1$ so
that $\psi$ is not simple. One then has (see (1.5))
$$x(\beta) = e_{-\psi} +
\sum_{i=1}^{\ell}\sqrt{n_i}\, e_{\alpha_i}\eqno (1.53)$$ and (see (1.30) and (1.27))
$$\overline {x(\beta)}= e_{\psi} + \sum_{i=1}^{\ell} \sqrt{n_i}\, e_{-\alpha_i}\eqno
(1.54)$$ Recalling (1.21) and (1.24) one notes that then $$\eqalign{\Re\,x(\beta)& =
(x(\beta) +\overline {x(\beta)})/2\cr
&= (e_{\psi} + e_{-\psi})/2  +
\sum_{i=1}^{\ell}\,\sqrt {\n_i}\,\,(e_{\alpha_i}+ e_{-\alpha_i})/2,\cr}\eqno (1.55)$$
and hence $$(\Re\,x(\beta),\Re\,x(\beta)) = h/2.\eqno (1.56)$$ But by (1.21) and (1.24)
one has $$\eqalign{\Im {x(\beta)}&= -i/2 ((x(\beta) -\overline {x(\beta)})\cr &=
-i/2((e_{\psi} - e_{-\psi})  +
\sum_{i=1}^{\ell}\,\sqrt {\n_i}\,\,(e_{\alpha_i}- e_{-\alpha_i})),\cr}\eqno
(1.57)$$ and hence $$(\Im\,x(\beta),\Im\,x(\beta)) = h/2.\eqno (1.58)$$ But clearly
(1.55) and (1.57) imply $$(\Re\,x(\beta),\Im\,x(\beta)) = 0.\eqno (1.59)$$ Let $Y\s
\hbox{Vec}\,\hh(\beta)$ be the two real-dimensional plane spanned by the
orthogonal vectors $\Re\,x(\beta)$ and $\Im\,x(\beta)$, and let
$$Q:\hbox{Vec}\,\,\hh(\beta)\to Y$$ be the orthogonal projection. Thus if $x\in
\hbox{Vec}\,\,\hh(\beta)$, then $$Qx = 2/h\,\,((x,\Re\,x(\beta))\,\Re\,x(\beta) +
(x,\Im\,x(\beta))\,\Im\,x(\beta)).\eqno (1.60)$$ But this implies $$(Qx,Qx) =
 2/h\,\,((x,\Re\,x(\beta))^2 + (x,\Im\,x(\beta))^2 ).$$ But now if $z = (x,x(\beta))$,
then $$\eqalign{(x,\Re\,x(\beta))&= \Re\,z\cr (x,\Im\,x(\beta))&= \Im\,z\cr}$$ by the
top lines in (1.55) and (1.57). Hence we have proved

\vs {\bf Proposition 1.18.} {\it
For any $x\in \hbox{Vec}\,\,\hh(\beta)$ one has $$|Qx|^2 = 2/h\,\,|(x,x(\beta))|^2.\eqno
(1.61)$$}\vs Now for any $\nu\in \Delta(\beta)$ (see \S 1.3) let $w_{\nu}\in
\hh(\beta)$ be defined, so that for any $x\in \hh(\beta)$, one has $\langle\nu,
x\rangle = (w_{\nu},x)$. Then as a consequence of (1.25) and Proposition 1.16 (where
now
$t_{\beta}=1$ by Theorem 1.12) and Proposition 1.18, one has

\vs {\bf Proposition 1.19.} {\it Let $\nu\in \Delta(\beta)$. Then $$|Qw_{\nu}|^2
={2\over h}\,\, |\nu_{\beta}|^2\eqno (1.62)$$} and our main result on the radius of
the two-dimensional orbit projections.

\vs {\bf Theorem 1.20.} {\it Let
$\beta\in
\Bbb C^{\times}$ be fixed so that $\beta = \beta^{(1)}$ is given as in Theorem 1.12.
Let $O_i,\,i=1,\ldots,\ell,$ be an orbit of the Coxeter
element $\sigma_{\beta}$ on the set $\Delta(\beta)$ of roots of $(\hh(\beta),\g)$.
Let $z_i$ be the corresponding basal element of $\hh$ defined as in (1.15). Then,
where $h$ is the Coxeter number, $z_i$ is an eigenvector of the operator (on $\hh$) $$
2/h\,\,\sum_{i=0}^{\ell}\, n_i\,w_{\alpha_i}\otimes w_{\alpha_i}\eqno (1.63)$$ and the
corresponding eigenvalue is $|Q\,w_{\nu}|^2$ where $\nu$ is any root in the orbit
$O_i$.}

\vskip 1.5pc\centerline{\bf 2. The special case of E$_8$}\vskip 1pc
{\bf 2.1.} Assume now that $\g$ is of type E$_8$. Then $\ell = 8$
and the cardinality of the set $\Delta$ of roots is 240. The Coxeter
number $h$ is 30. The group is unique up to isomorphism. In
particular $G\cong G_{ad}$.  Let $\beta\in (\Bbb C^{\times})^9$ be
as in Theorem 1. The Gosset polytope (see e.g., [Go]) published in
1900 may be taken to be the boundary of the convex hull of the
vectors $w_{\gamma}, \, \gamma\in \Delta(\beta)$, in the
8-dimensional real space $\hbox{Vec}\,\hh(\beta)$. The Coxeter
element $\sigma_{\beta}$ decomposes $\Delta(\beta)$ into 8 orbits
$O_i,\,i=1,\ldots 8,$ where each orbit contains 30 roots. Peter
McMullen made a drawing of a two real-dimensional projection of the
Gosset polytope. It appears as the frontispiece of Coxeter's book
[CX]. The projection is now quite famous and appears in many places
in the literature. The image of the orbits in
$\hbox{Vec}\,\hh(\beta)$ corresponding to the $O_i$ appears as 8
concentric circles, which, by abuse of notation, we will refer to as
the Gosset circles. Our main objective here is to determine the
ratio of the radii of the Gosset circles. That Theorem 1.20
accomplishes this is a consequence of John Conway's identification
of McMullen's projection with the map $Q$.

\vs {\bf Remark }. One is forced into Conway's identification if one
demands that the projection commutes with the action of the Coxeter
element. Indeed since in the E$_8$ case all the 8 eigenvalues of
$\sigma_{\beta}$ are primitive 30th roots of unity,  the
corresponding eigenvectors are cyclic elements and hence are Weyl
group conjugate by elements which normalize the cyclic group
generated by $\sigma_{\beta}$. It follows that there are only 4
two-dimensional real projections which commute with the action of
the Coxeter element $\sigma_{\beta}$ and all four are isomorphic to
$Q$.

\vs {\bf 2.2.} {\bf Remark. } As one knows the E$_8$ root lattice
can be constructed from the golden number and the embedding of the
120 element binary icosahedral group in the group of unit
quaternions. It therefore may be more than a coincidence to note
that the $n_i$ appearing in the construction of $A$ are, by the
McKay correspondence, the dimensions of the irreducible
representations of the binary icosahedral group.

\vs David Vogan reexpressed the operator
$A$ as an element
$A'$ in the group algebra of the Weyl group. Letting $F$ be the characteristic polynomial
of a convenient multiple of $A'$, he found that $F$ factors into a product of 2
irreducible (over $\Bbb Q$) degree 4 polynomials $F_1$ and $F_2$, where
$$\eqalign{F_1(x)&= x^4- 15x^3 +75 x^2 -135 x + 45\cr F_2(x)&= x^4 -15 x^3 +60 x^2 -90 x
+45 .\cr}\eqno (2.1)$$ Vogan then computed the integral part of the radii of the Gosset
circles normalized so that the maximal integral part is 1000. They are, in increasing size,
$$\matrix{209\cr 338\cr 416\cr 502\cr 618\cr
673\cr 813\cr 1000\cr}\eqno (2.2)$$ The use of quotation marks in the following statements is a
consequence of the statement in the Remark of \S 0.3. \vskip .5pc {\bf ``Theorem" 2.1.} {\it The ratio of the
normalized radii in (2.2) ``agrees" with the ratio of the conjectured 8 masses in [Za]. See (1.8) in [Za].}\vs

We later found out that the ratio of the smallest Gosset circles (the larger over the smaller) should be the
Golden number $R = {1\over 2}(1 +
\sqrt 5)$.
 Finding this to be the case experimentally was the key discovery in [Co].
        The decomposition $F= F_1 F_2$ implies that the set of Gosset circles decomposes into two
sets of 4 Gosset circles. The radii of one set can be expressed in terms of the radii of
the other set using $R$ and $1/R$ as follows:\vskip .5pc
$$\matrix{209&\times&R&``="&338\cr 673&\times&1/R&``="&416\cr
813&\times&1/R&``="&502\cr 618&\times&R&``="&1000\cr}\eqno (2.30)$$\vskip .5pc
Here the first column is filled with the normalized radii of the Gosset circles defined by
$F_1$ and the last column is filled with the normalized radii of the Gosset circles defined by
$F_2$.
 \vskip 1pc \centerline{\bf References}\vskip 3pt
\item {[Ba]} J. Baez, {\it Week 289}, Jan 8, 2010

\item{[Co]} R. Coldea, D.A.Tennant, E.M. Wheeler, E. Wawrzynska, D. Prabhakaran, M. Telling, K. Habnicht, P. Smeibidl, K. Kiefer, Quantum Criticality in
an Ising Chain: Experimental Evidence for Emergent, E$_8$ Symmetry,
{\it Science} {\bf 327}, 8 January 2010,  177--180

\item{[CX]}  H.S.M. Coxeter, {\it Regular Complex Polytopes}, Cambridge Univ. Press, 1974

\item {[Go]} T. Gosset, On the regular and semi-regular figures in space of $n$ dimensions,
{\it Messenger of Mathematics}, {\bf 29} (1900), 43--49

\item {[K-59]} B. Kostant, The Three-Dimensional Sub-Group and the Betti Numbers of a Complex Simple Lie
Group, {\it Amer. Jour. of Math.}, {\bf 81}(1959), 973--1032

\item{[Za]} A.B. Zamolodchikov, Integrals of Motion and S-matrix of the (Scaled) $T=T_c$ Ising Model with
Magnetic Field, {\it International Journal of Modern Physics}, {\bf 4} No.~16 (1989), 4235--4248

\vskip 1pc
\noindent Bertram Kostant

\noindent Department of Mathematics (Emeritus)

\noindent  MIT

\noindent Cambridge, MA 02139

\noindent email: kostant@math.mit.edu

\end